\documentclass{PoS}
\usepackage{amssymb,amsbsy,amsmath,amsfonts}
\usepackage{mathrsfs}
\usepackage{graphicx}% Include figure files
\usepackage{graphics,psfrag,longtable}
\usepackage{nicefrac}
\usepackage{esint}
\usepackage{upgreek}
\DeclareMathAlphabet{\mathantt}{OML}{antt}{l}{it}
\DeclareMathAlphabet{\mathpzc}{OT1}{pzc}{m}{n}

\def\beq{\begin{equation}}
\def\eeq{\end{equation}}
\def\bea{\begin{eqnarray}}
\def\eea{\end{eqnarray}}
\def\beqa{\begin{equation}\begin{array}{l}}
\def\eeqa{\end{array}\end{equation}}
\def\eqlab#1{\label{eq:#1}}

\def\eref#1{(\ref{eq:#1})}
\def\Eqref#1{Eq.~(\ref{eq:#1})}

\def\barr{\left(\begin{array}{c}}
\def\earr{\end{array}\right)}
\def\bmat{\left(\begin{array}{cc}}
\def\emat{\end{array}\right)}
\def\al{\alpha}

\def\ga{\gamma}

 \def\La{{\Lambda}}

\def\cA{\mathcal{A}}
\def\bQ{\mathbf{Q}}
\def\br{\mathbf{r}}

\def\nn{\nonumber}
\DeclareMathOperator\arccosh{arccosh}
\def\dd{\mathrm{d}}

\DeclareMathOperator\im{Im}
\def\3d{3-D}

\def\bq{\mathbf{q}}

\title{Proton structure in the hyperfine splitting of muonic hydrogen}

\ShortTitle{Proton structure in the hyperfine splitting of muonic hydrogen}

 \author{\speaker{Franziska Hagelstein}\thanks{This work was supported by the Deutsche Forschungsgemeinschaft (DFG) through the Collaborative Research Center SFB 1044 [The Low-Energy Frontier of the Standard Model], and the Graduate School DFG/GRK 1581 [Symmetry Breaking in Fundamental Interactions].} \hspace{1mm}and Vladimir Pascalutsa\\
 Institut für Kernphysik, Johannes Gutenberg Universität, Mainz, Germany \\
     E-mail:  \email{hagelste@uni-mainz.de}}

\abstract{We present the leading-order prediction of  baryon chiral perturbation theory for the proton polarizability contribution to the $2S$ hyperfine splitting in muonic hydrogen, and compare 
to the results of dispersive calculations.  }

\FullConference{The 8th  International Workshop on Chiral Dynamics \\
		 29 June 2015 - 03 July 2015\\
		 Pisa, Italy}
\begin{document}
\section{Introduction}
%The apparent discrepancy between extractions of the proton radius from experiments performed with electronic \cite{Mohr:2012tt,Bernauer:2010wm,Zhan:2011ji} vs.\ muonic probes \cite{Antognini:1900ns,Antognini:2013rsa} -- {\it aka} the proton-size puzzle -- calls for a precise theoretical understanding of the hydrogen spectrum. 
The  extraction of the proton radius from muonic hydrogen ($\mu$H) spectroscopy relies on a comparison between measured transition frequencies in the hydrogen atom \cite{Antognini:1900ns} and theoretical predictions for the hydrogen spectrum. The description of the classic $2P-2S$ Lamb shift and the $2S$ hyperfine splitting (HFS) in $\mu$H is given by (in units of meV)~\cite{Antognini:2013rsa}:
\begin{subequations}
\eqlab{Htheory}
\bea
\Delta E_{\mathrm{LS}}^{\mathrm{th}}=206.0336(15)-5.2275(10)\,(R_E/\mathrm{fm})^2+\Delta E^{\mathrm{TPE}}_{\mathrm{LS}}\, , \quad &\text{with}& \,\, \Delta E^{\mathrm{TPE}}_{\mathrm{LS}}=0.0332(20),\qquad\eqlab{LS theory}\\
%&=&206.0668(25)-5.2275(10)\,(R_E/\mathrm{fm})^2,\nn\\
\Delta E_{2S\,\mathrm{HFS}}^{\mathrm{th}}=22.9763(15)-0.1621(10)\,(R_Z/\mathrm{fm}) +\Delta E^{\mathrm{pol}}_{2S\,\mathrm{HFS}}\, , \quad &\text{with}& \,\,\Delta E^{\mathrm{pol}}_{2S\,\mathrm{HFS}}=0.0080(26),\hspace{1cm}
\eqlab{HFStheory}
%&=&22.9843(30)-0.1621(10)\,(R_Z/\mathrm{fm}),\nn
\eea
\end{subequations}
where $R_E$ is the proton charge radius, $R_Z$ is the Zemach radius, and $\Delta E^{\mathrm{TPE}}_{\mathrm{LS}}$ together with $\Delta E^{\mathrm{pol}}_{2S\,\mathrm{HFS}}$ are proton structure effects beyond leading order (LO). We investigate both the finite-size and the polarizability effects on the HFS.

\section{Finite-size effects by dispersive technique}
\begin{figure}[tbh]
\vspace{1mm}
\centering
\includegraphics[scale=0.5]{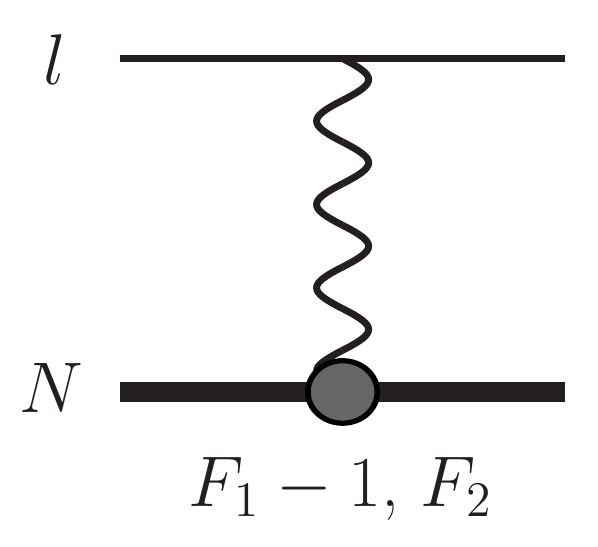}
    \caption{One-photon exchange with form factor dependent electromagnetic vertex. Here, $l$ represents the lepton and $N$ the nucleon.\label{fig:1gamma}}  
\end{figure}
In a recent paper \cite{Hagelstein:2015yma}, we study the applicability of the expansion of finite-size corrections to the Lamb shift in moments of the charge distribution. We now present an extension of this work to the HFS.

Figure \ref{fig:1gamma} shows the one-photon exchange in an atomic bound state. It can be calculated using the electromagnetic vertex:
\beq
\Gamma^\mu =Z \ga^{\,\mu} F_1(Q^2) -\frac{1}{2M} \ga^{\,\mu\nu} q_\nu
F_2(Q^2),\quad (Z=1 \; \text{for the proton}) \eqlab{protonphotonvertex}
\eeq
with the proton structure information embedded in the Dirac and Pauli form factors, $F_1$ and $F_2$, and the photon propagator in massive Coulomb gauge:
\beq
\Delta_{\mu\nu}(q,t)=-\frac{1}{q^2}\left[g_{\mu\nu}-\frac{1}{\bq^2+t}\left(q_\mu q_\nu-\chi_\mu q_\nu-\chi_\nu q_\mu\right)\right], \quad \text{with } \,\,\chi=(0,\bq)\eqlab{photonProp}.
\eeq
Assuming that the nucleon form factors fulfill once-subtracted dispersion relations,
\bea
\barr F_1 (Q^2)\\ 
F_2 (Q^2) \earr  = \barr 1\\
\varkappa \earr + \frac{q^2}{\pi} \int_{t_0}^\infty \frac{\dd t}{ t (t-q^2) }
\im \barr  F_1(t) \\ 
F_2 (t) \earr, \eqlab{F12DR}
\eea
with $t_0$ being the lowest particle-production threshold and $q^2=-Q^2$ the virtuality of the exchanged photon, we can deduce the following one-photon exchange (coordinate space) Breit potential:
\begin{subequations}
\bea
V_\mathrm{F.S.}(\br,t)&=&\frac{Z \al}{\pi r}\int_{t_0}^\infty\frac{\dd t}{t} e^{-r\sqrt{t}} \im G_E(t),\eqlab{Yukawa}\\
V^{l=0}_{\mathrm{ HFS, a}}(\br,t)& = &\frac{Z\al}{3}\frac{1+\varkappa}{mM}\left[f(f+1)-\frac{3}{2}\right]\,4\pi\,\delta(\br),\eqlab{HFSa}\\
 V^{l=0}_{\mathrm{ HFS, b}}(\br,t)& = &-\frac{4Z\al}{3}\frac{1+\varkappa}{mM}\left[f(f+1)-\frac{3}{2}\right]\left( \,\delta(\br)\int_{t_0}^\infty\frac{\dd t}{t}\frac{\im G_M(t)}{1+\varkappa}-\pi \rho_M(r)\right),\eqlab{HFSb} \\
% & = & -\frac{Z\al}{3}\frac{1+\varkappa}{mM}\left[f(f+1)-\frac{3}{2}\right]\frac{1}{\pi}\int_{t_0}^\infty\frac{\dd t}{t}\left(4\pi \,\delta(\br)-\frac{t e^{-r\sqrt{t}}}{r}\right)\frac{\im G_M(t)}{1+\varkappa},\eqlab{HFSb}\qquad \\
 \vdots&&\nn
\eea
\end{subequations}
where we substituted the electric and magnetic Sachs form factors:
\bea 
&& G_E(Q^2)  = F_1(Q^2) - \tau F_2(Q^2), \quad G_M(Q^2) = F_1 (Q^2)+ F_2(Q^2),
\eea
with $\tau = Q^2/4M^2$. Hereinafter, $m$ refers to the muon mass, $M$ is the proton mass, $\varkappa$ is the anomalous magnetic moment of the proton, $l$ is the orbital angular momentum ($l=0$ for $S$-waves), $f$ is the atom's total angular momentum and $\rho_M$ is the magnetization density:
 \bea
 \rho_M (r) &=&\int\! \frac{\dd\bQ}{(2\pi)^3} \frac{G_M(Q^2)}{1+\varkappa} e^{i \bQ\cdot \br} = \frac{1}{(2\pi)^2\, r} \int^\infty_{t_0} \! \! \dd t\, \frac{\im G_M(t)}{1+\varkappa}\,  e^{-r\sqrt{t}} .
 \eea

At $1^\mathrm{st}$-order in time-independent perturbation theory (PT), the Lamb shift due to the Yukawa-type correction in \Eqref{Yukawa} is given by:
\begin{subequations}
 \bea
 E^{\mathrm{FSE}\,(1)}_\mathrm{LS} & =& -\frac{Z\al}{2\pi a^3} \int_{t_0}^\infty \!\!\dd t \, \frac{\im G_E (t)}{(\sqrt{t}+Z\al m_r)^4}, 
\eqlab{rmsLSa}\\
&=& -\frac{Z\al}{12 a^3} \sum_{k=0}^\infty 
 \frac{(-Z\al m_r)^{k}}{k!} \langle r^{k+2}\rangle_E,\quad\eqlab{rmsLSaEXP}
\eea
\end{subequations}
or equivalently,
  \beq
 E^{\mathrm{FSE}\,(1)}_\mathrm{LS}=\int_0^\infty \dd Q \, w_E(Q)\,G_E(Q^2), \quad \text{with }\,\, w_E(Q)=-\frac{4 Z\al}{\pi a^4} 
\frac{Q^2\left[ (Z\al m_r)^2-Q^2\right]}{\left[(Z\al m_r)^2+Q^2\right]^4}, \eqlab{LSexact}\\
%&=& -\frac{Z\al \pi}{3a^3}  \int_0^\infty\! \dd r \, r^4 e^{-r/a } \rho_E (r)
  \eeq
%\end{subequations}
where $m_r$ is the reduced mass of the muon-proton system and $a=(Z\al m_r)^{-1}$ is the Bohr radius.
In deriving \Eqref{rmsLSaEXP}, we have expanded in the moments of the charge distribution, using the following (Lorentz-invariant) definition:
\beq
\eqlab{rmsdef}
\langle r^N\rangle_E =
\frac{(N+1)!}{\pi}\int_{t_0}^\infty\!\! \dd t \, \frac{\im  G_E(t)}{t^{N/2+1 }  },\quad \text{where}\;\; R_E =  \sqrt{\langle r^2\rangle_E}.
\eeq
\begin{figure}[htb]
\centering
\includegraphics[scale=0.56]{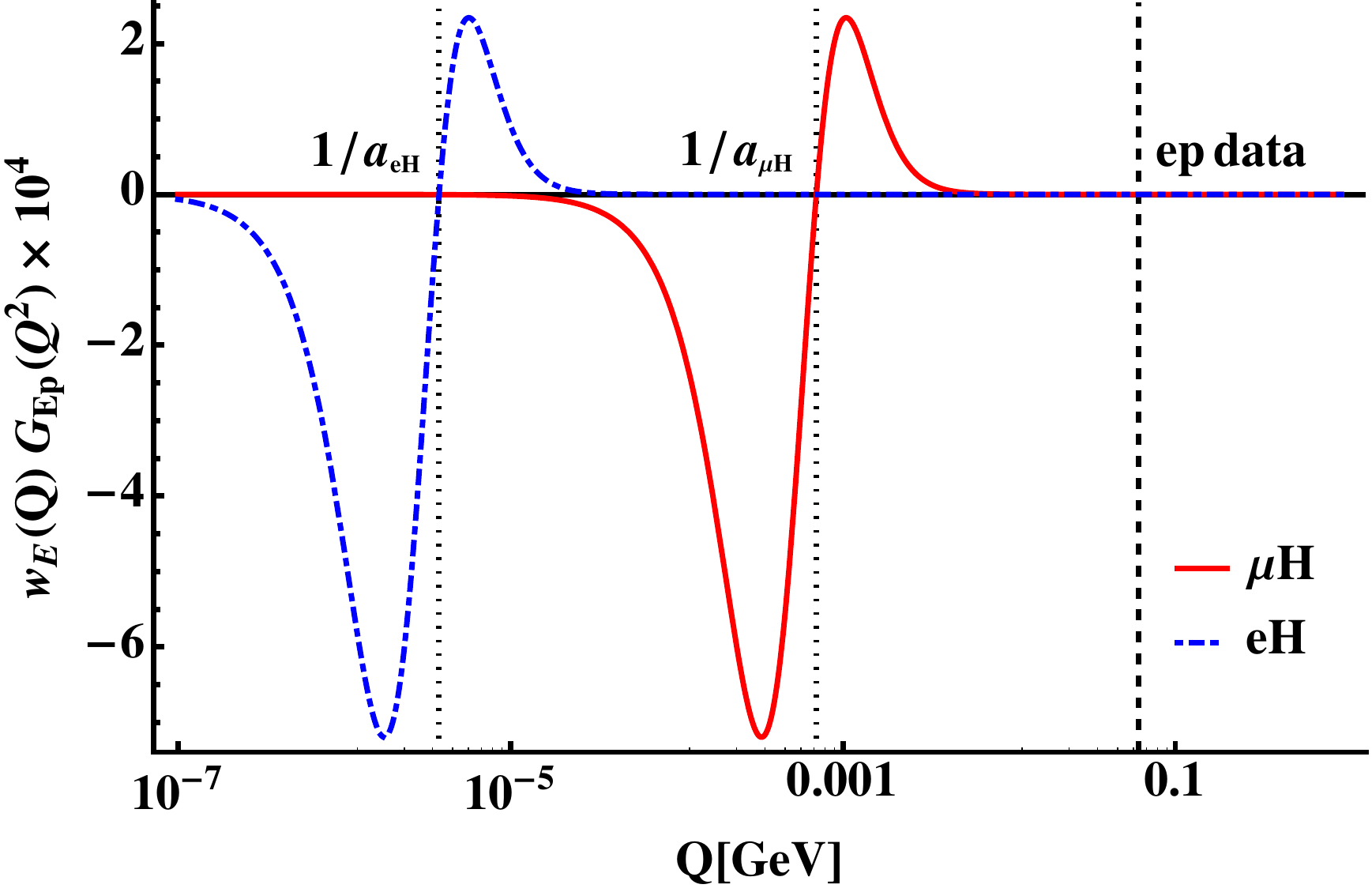}
\caption{
Integrand of the 1\textsuperscript{st}-order 
PT contribution to the Lamb shift [cf.\ Eq.~(2.8)]
       of $e$H (blue dash-dotted line) and of $\mu$H (red solid line) for the dipole FF, 
       $G_{Ep}= (1+Q^2/0.71\,\mathrm{GeV}^{\,2})^{-2}$. The dotted vertical lines
       indicate the inverse Bohr radii of the two hydrogens, while the vertical dashed line indicates the onset of the elastic
       electron-proton scattering data.}
       \label{fig:IntegrandLS}
\end{figure}
However, from the denominator of \Eqref{rmsLSa} one can see that the convergence radius of the power-series expansion is limited
by $t_0$, i.e., the proximity of the nearest particle-production threshold. The exact (non-expanded) finite-size effect on the Lamb shift, cf.\ \Eqref{LSexact}, is the result of large cancelations around the Bohr radius scale, see Fig.~\ref{fig:IntegrandLS}.
We conclude that soft contributions to the proton or lepton electric form factor, at energies comparable to the inverse Bohr radius, can break down the expansion, and accordingly, limit the usual accounting of finite-size effects which is presented in \Eqref{LS theory}. Therefore, one has to know all the soft contributions to the proton electric form factor to high accuracy for an accurate extraction of $R_E$ from the Lamb shift in $\mu$H.

\begin{figure}[tbh]
\centering
 \includegraphics[scale=0.56]{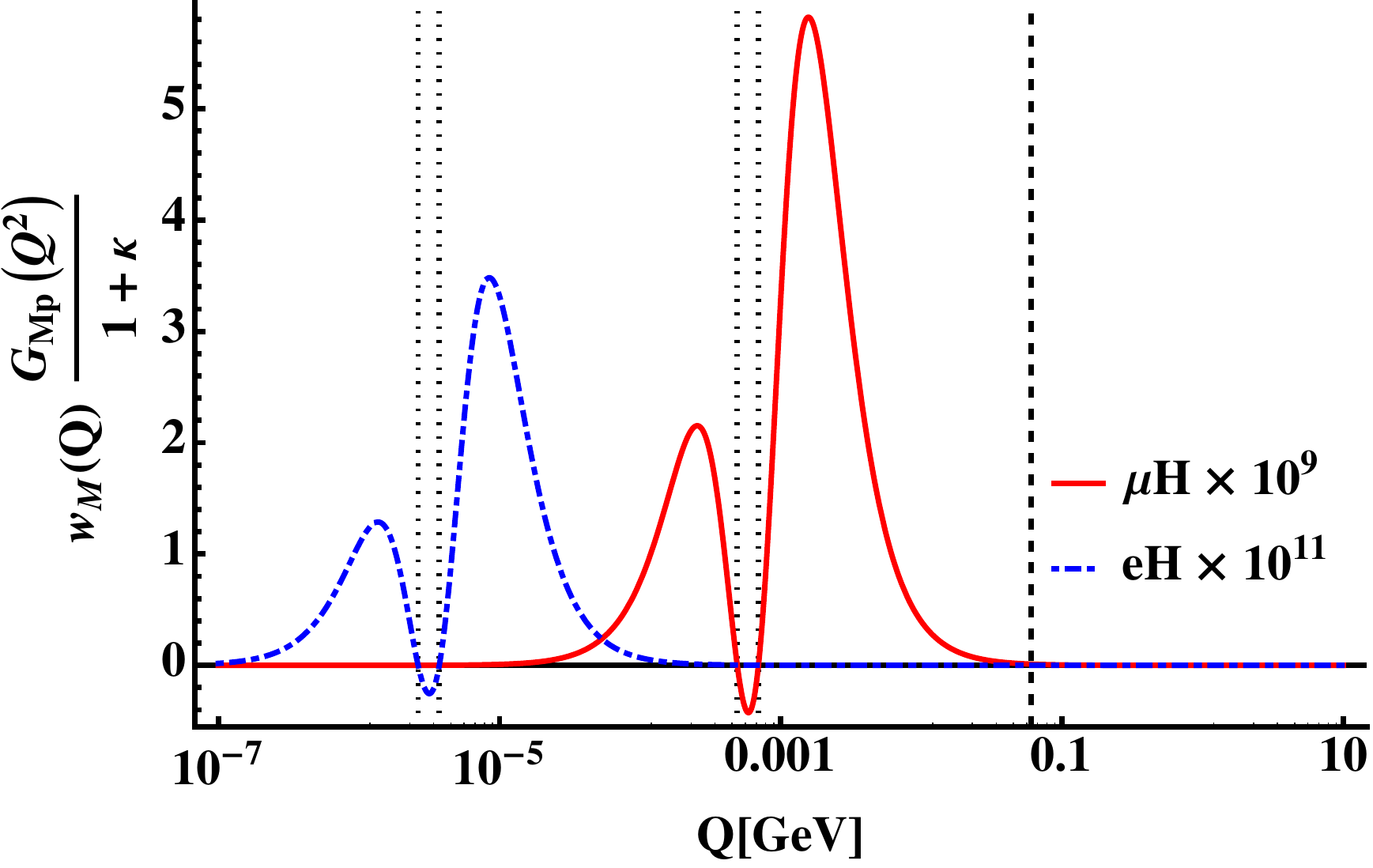}
\caption{Integrand of the $1^\mathrm{st}$-order PT contribution to the $2S$ HFS [cf.\ Eqs.~(2.10) and (2.11b)], where $w_M(Q)=\frac{4E_F(2S)}{\pi a} \frac{Q^2\left[ (Z\al m_r)^2-Q^2\right]\left[ (Z\al m_r)^2-2Q^2\right]}{\left[(Z\al m_r)^2+Q^2\right]^4}$, of $e$H (blue dash-dotted line) and of $\mu$H (red solid line) for the dipole FF, $G_{Mp}= (1+\varkappa)(1+Q^2/0.71\,\mathrm{GeV}^{\,2})^{-2}$. The dotted vertical lines indicate scales related to the inverse Bohr radii of the two hydrogens, particularly $\nicefrac{1}{a}\,$ and $\nicefrac{1}{\sqrt{2}a}\,,$ while the vertical dashed line indicates the onset of the elastic electron-proton scattering data.}
\label{fig:IntegrandHFS}
\end{figure}

Analogously, we establish the exact finite-size effect on the HFS, starting from the spin-dependent part of the Breit potential, cf.\ Eqs.~\eref{HFSa} and \eref{HFSb}. Treating the potential \eref{HFSa} at $1^\mathrm{st}$-order in PT, we reproduce the LO HFS, which is given by the well-known Fermi energy:
\beq
E_F(nS)=\frac{8 Z\al}{3a^3}\frac{1+\varkappa}{mM} \frac{1}{n^3}\overset{n=2}{=}22.8054\, \mathrm{ meV}.\eqlab{HFS1}
\eeq
Applying $1^\mathrm{st}$-order PT to \Eqref{HFSb} leads to:
\begin{subequations}
\eqlab{HFS1PT}
 \bea
 E^{\mathrm{HFS, b}\,(1)}_{2S\, \mathrm{ HFS}} & =& -\frac{E_F(2S)}{\pi}  \int_{t_0}^\infty \!\!\dd t \, \left(\frac{1}{t}-\frac{2t+(Z\al m_r)^2}{2[\sqrt{t}+Z\al m_r]^4}\right)  \frac{\im G_M (t)}{1+\varkappa},
\eqlab{HFS11}\\
&=& -E_F(2S)\left\{1-\frac{4}{\pi a} \int_0^\infty \dd Q \frac{Q^2\left[ (Z\al m_r)^2-Q^2\right]\left[ (Z\al m_r)^2-2Q^2\right]}{\left[(Z\al m_r)^2+Q^2\right]^4}\frac{G_M(Q^2)}{1+\varkappa}\right\},\qquad\quad\eqlab{HFSexact}\\
&=& -E_F(2S)\left\{1- 8\pi a^3 \int_0^\infty \dd r\, r^2 \big[R_{20}(r)\big]^2 \,\rho_M(r)\right\},\eqlab{HFS1PTc}
\eea
\end{subequations}
with the radial Coulomb wave function  $R_{20} (r) = \nicefrac{1}{\sqrt{2} \, a^{3/2}} \left(1- \nicefrac{r}{2a} \right) 
 e^{-r/2a}$.
Reading off Eqs.~\eref{HFS1PTc} and \eref{HFS1}, the combined $1^\mathrm{st}$-order result is entirely expressed as an integral over the magnetization density. The corresponding integrand is plotted in Fig.~\ref{fig:IntegrandHFS} for the dipole FF. Again, one observes an enhancement of the integrand for small values of $Q$, which emphasizes the necessity for an exact evaluation of soft contributions to the form factors.

\section{Polarizability contribution to the HFS}
In this section, we give a baryon chiral perturbation theory (BChPT) prediction for the pion contribution to the polarizability effect on the $2S$ HFS in $\mu$H, cf.\ $\Delta E^{\mathrm{pol}}_{2S\,\mathrm{HFS}}$ from \Eqref{HFStheory}. 

\begin{figure}[hb]
\centering
\includegraphics[width=6cm]{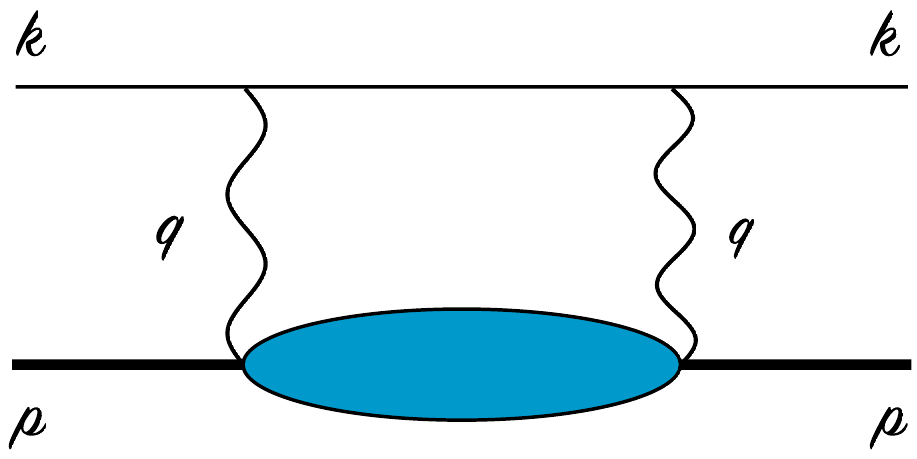}
\caption{TPE diagram in forward kinematics: the horizontal lines correspond to the lepton and the proton (bold), where the ``blob'' represents forward doubly-virtual Compton scattering.}
\label{TPEFRW}
\end{figure}
Figure~\ref{TPEFRW} shows the (forward) two-photon exchange (TPE) in an atomic bound state, which contributes to the HFS and the Lamb shift at next-to-leading order. Fading out the lepton line in Fig.~\ref{TPEFRW}, one obtains the doubly-virtual Compton scattering process (VVCS). For the required accuracy, $O(\al^5)$, it is sufficient to study only the forward limit, i.e., $q'=q$ (hence, $p'=p$, $t=0$). The tensor decomposition of the forward VVCS amplitude then splits into symmetric
and antisymmetric parts:
\begin{subequations}
\begin{eqnarray}
T^{\mu \nu}(q,p) &=& T^{\mu \nu}_S+T^{\mu \nu}_A,\\
T^{\mu \nu}_S(q,p) & = & -g^{\mu\nu}
T_1(\nu, Q^2)  +\frac{p^{\mu} p^{\nu} }{M^2}  T_2(\nu, Q^2), \label{eq:VVCS_TS}\\
T^{\mu \nu}_A (q,p) & = &\frac{1}{M}\gamma^{\,\mu \nu \al} q_\al S_1(\nu, Q^2) -
\frac{Q^2}{M^2}  \gamma^{\,\mu\nu} S_2(\nu, Q^2),
\end{eqnarray}
\end{subequations}
with $T_{1,2}$ the spin-independent and $S_{1,2}$ the spin-dependent invariant amplitudes,
functions of the photon (lab-frame) energy $\nu$ and the photon virtuality $Q^2=-q^2$. The spin-dependent part of the Compton process contributes to the HFS, whereas the spin-independent part contributes to the Lamb shift. 
The TPE correction to the HFS of the $n$-th $S$-level is given by \cite{Carlson:2011af}:
\bea
E_{nS \,\mathrm{HFS}}^\mathrm{pol}&=&\frac{m\, E_F(nS)}{2(1+\varkappa)\pi^4}\frac{1}{i}\int_0^\infty \dd \nu \int \dd \bq\,\frac{1}{Q^4-4m^2\nu^2}\left\{\frac{\left(2Q^2-\nu^2\right)}{Q^2}S_1(\nu,Q^2)+3\frac{\nu}{M}S_2(\nu,Q^2)\right\}\eqlab{VVCS_HFS}, \qquad\,
\eea
where $E_F$ is the hydrogen Fermi energy, i.e., the LO HFS. The TPE can be divided into an ``elastic'' and an ``inelastic'' part. As mentioned previously, we are only interested in the ``polarizability'', or ``inelastic'', contribution given by the non-Born part of the Compton amplitudes.

All invariant amplitudes are related to photoabsorption cross sections by sum rules, however, the amplitude $T_1$ requires a once-subtracted dispersion relation. Therefore, in contrast to the HFS, the polarizability contribution to the Lamb shift
is not determined by empirical information (on nucleon form factors, and structure functions or photoabsorption cross sections) alone and requires a rigorous theoretical input. 
Such an input has been provided by the recent BChPT calculation of Alarcón et al.\ \cite{Alarcon:2013cba}. We extend this
calculation to the HFS, where the BChPT framework is put to the test by the available dispersive calculations \cite{Carlson:2011af,Carlson:2008ke,Faustov:2006ve,Faustov:2001pn,Martynenko:2004bt}. Moreover, we address the contributions to the HFS by pion exchange, which are off-forward and not covered by  \Eqref{VVCS_HFS}.\subsection{BChPT at leading order}
\begin{figure}[hb]
\centering
\includegraphics[width=10cm]{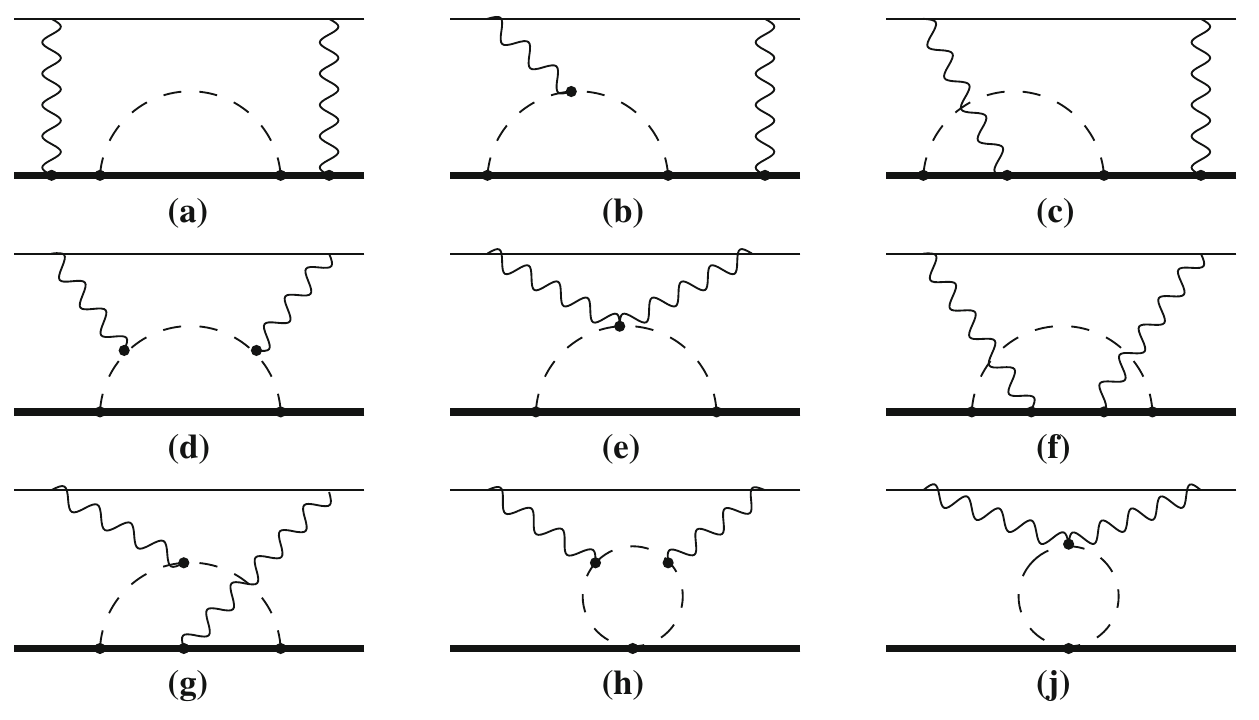}
\caption{The TPE diagrams of elastic lepton-nucleon scattering to $O(p^3)$ in BChPT. Diagrams obtained from these by crossing and time-reversal symmetry are not drawn. Reproduced from Ref.~\cite{Alarcon:2013cba}.}
\label{PiNLoops}
\end{figure}
\begin{figure}[!h]
\centering
\includegraphics[scale=0.5]{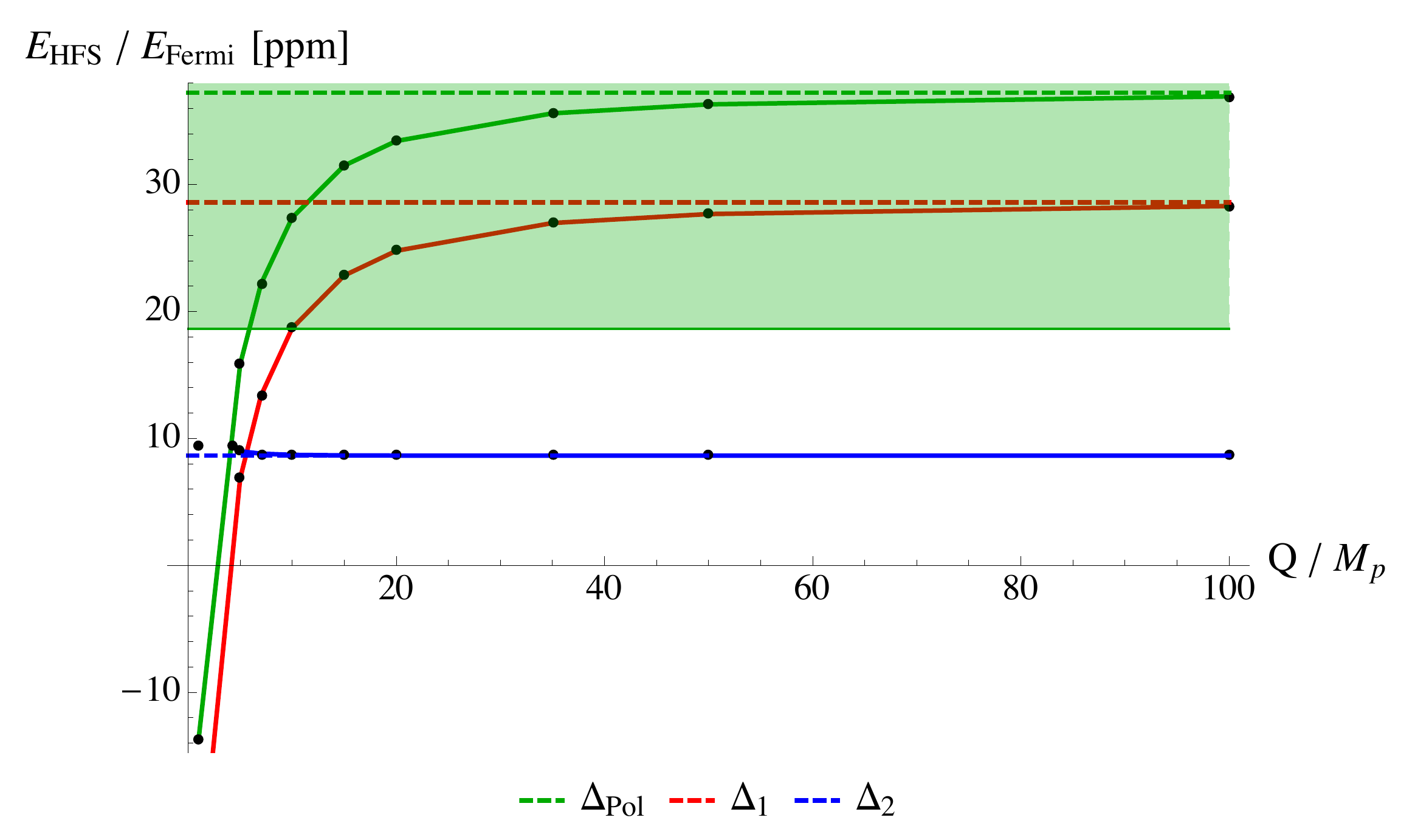}
\caption{Cutoff-dependence of the pion-nucleon loop contribution. Here, $\Delta_1$ and $\Delta_2$ represent the contributions due to $S_1$ and $S_2$, respectively. $\Delta_\mathrm{pol}$ is the sum of $\Delta_1$ and $\Delta_2$. The dashed lines are without momentum-cutoff.}
\label{PiNCut}
\end{figure}

The pion-nucleon loop contribution to the HFS, shown in Fig.~\ref{PiNLoops}, is evaluated based on \Eqref{VVCS_HFS}. Substituting the Compton amplitudes from Ref.~\cite{Alarcon:2013cba}, we obtain the $2S$ HFS in $\mu$H:
\bea
E_{2S\,\mathrm{HFS}}^{(\pi N\,\mathrm{loops})}=0.85\pm 0.42\, \upmu \mathrm{eV}.
\eqlab{PNLoop}
\eea
This is the pure ``polarizability''
contribution, i.e., corrections due to elastic form factors have been subtracted. Figure \ref{PiNCut} shows the dependence of the pion-nucleon loop polarizability contribution (green line) on a momentum-cutoff. We estimate the error with $50\,\%$, what is illustrated by the green band.

\subsection{Remarks on $\pi^0$ exchange}
\begin{figure}[!h]
\centering
\includegraphics[scale=0.5]{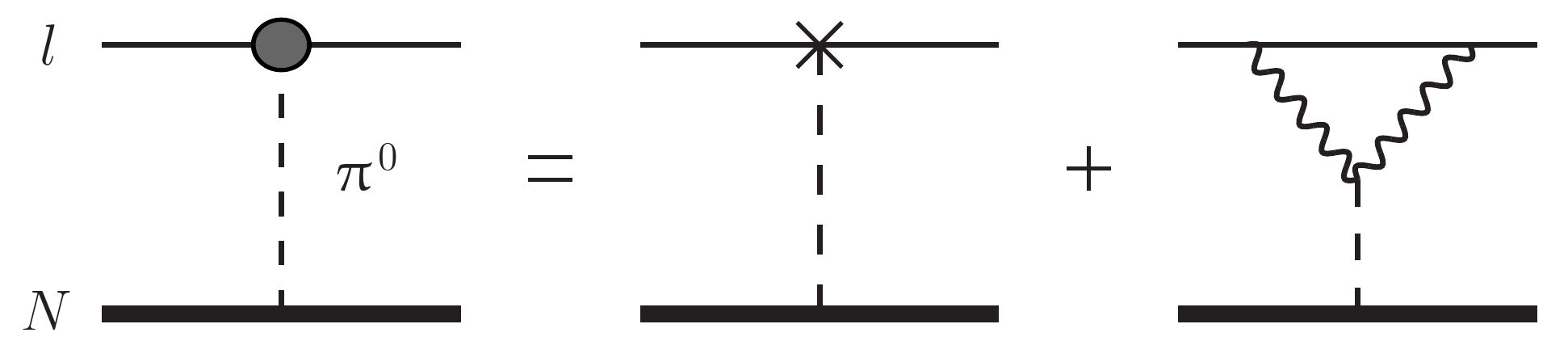}
\caption{$\pi^0$ exchange in an atomic bound state. Here, $l$ represents the lepton and $N$ the nucleon.}
\label{PiExchange}
\end{figure}
Figure \ref{PiExchange} shows the neutral-pion exchange in an atomic bound state. This process is vanishing in the forward limit, but gives a $O(\al^6)$ contribution for off-forward scattering. For the HFS contribution we obtain:
\beq
E_{nS\,\mathrm{HFS}}^{(\pi^0)}= - E_F(nS)
\frac{\al^2 }{2\pi(1+\varkappa_p) }\frac{m_r}{m_{\pi}}g_{\pi NN}\left[g_{\pi \ell\ell}+\frac{m}{2\pi^2 f_\pi}I(\kappa)\right],
 \eeq
 where $\kappa=m_\pi/2m$, and
 \beq
 I(\kappa)\equiv 2 \int_0^\infty \frac{\dd \zeta}{1+\zeta/\kappa}\frac{\arccos \zeta}{\sqrt{1-\zeta^2}}.
 \eeq
Note that the first term is due to the diagram with the pion directly coupling to the lepton and the second term is due to the diagram with two-photon coupling to the lepton. The $\pi \ell \ell$ coupling can be extract from the decay width of $\pi^0\to e^+ e^-$:
\beq
\Gamma(\pi^0\to e^+ e^-) = \frac{m_\pi}{8\pi} \sqrt{1-\frac{4m_e^2}{m_\pi^2}} \, \big| F(m_\pi^2,m_e^2,m_e^2) \big|^2,
\eeq
with $m_\pi$ and $m_e$ being the mass of pion and electron, respectively. Since the $\pi \ell \ell$ coupling is given by: 
\beq
g_{\pi\ell\ell}=F(0,m_\ell^2,m_\ell^2)/ \al^2,
\eeq 
an interpolation between the pion momenta $q^2 = 0$ and $m_\pi^2$ is required. To leading order in $\al$, the form factor might be written as:
\begin{subequations}
\bea
 F(q^2) \equiv  F(q^2,m_\ell^2,m_\ell^2)  &=& F(0) + \frac{q^2}{\pi} \int_0^\infty \!\frac{\dd s}{s}\,  \frac{\im F(s)}{s-q^2}  , \\
  \im F(s) &=& -\frac{\al^2 m_\ell}{2\pi f_\pi} \frac{\arccosh(\sqrt{s}/2m_\ell)}{\sqrt{1-4m_\ell^2/s}}, \\
 F(0)  &=&  \frac{\al^2 m_\ell}{2\pi^2 f_\pi} \Big( \cA (\La) + 3\ln\frac{m_\ell}{\La} \Big),
\eea 
\end{subequations}
where $f_\pi$ is the pion-decay constant, $\La$ is the renormalization scale, and $\cA$ is a universal pion-lepton low-energy constant,
related to the physical constant in an obvious way:
\beq
g_{\pi\ell\ell} = \frac{m_\ell}{2\pi^2 f_\pi} \cA (m_\ell). 
\eeq
From the experimental $\pi^0\to e^+ e^-$ decay width one finds: $\cA (m_e) = -20(1)$.  The value of the muon coupling
we then find as: $\cA (m) =\cA (m_e) + 3\ln (m/m_e )=-4(1)$, with the muon mass $m$.
The $\pi NN$ coupling can be approximated by the Goldberger-Treiman
relation:
\beq
g_{\pi NN} = M g_A/f_\pi,
\eeq
with the axial coupling $g_A\approx 1.27$ and $f_\pi\approx 92.4 \,\mathrm{MeV}$.
Finally, the HFS of the $2S$ level in $\mu$H amounts to:
%\bea
%E_\mathrm{HFS}^{(\pi^0)}=0.44\pm 0.04\, \upmu\text{eV}.
%\eea
\bea
E_{2S\,\mathrm{HFS}}^{(\pi^0)}=0.02\pm 0.04\, \upmu\mathrm{eV}.
\eqlab{PE}
\eea

Recently, the effect of the pion exchange has also been estimated by Zhou et al.\ \cite{Zhou:2015bea}. They find the diagram with two-photon coupling (cf. Fig.~\ref{PiExchange}) to dominate, and arrive at a two orders of magnitude larger contribution: $E_{2S\,\mathrm{HFS}}^{(\pi^0)}=2.8\, \upmu\mathrm{eV}$. However, they simply evaluated the ultraviolet-divergent loop in Fig.~\ref{PiExchange} by imposing an arbitrary cutoff and did not check whether their result is consistent with the $\pi^0\to e^+ e^-$ decay width.

\section{Discussion and conclusion}
\begin{figure}[thb]
\centering
\includegraphics[scale=0.8]{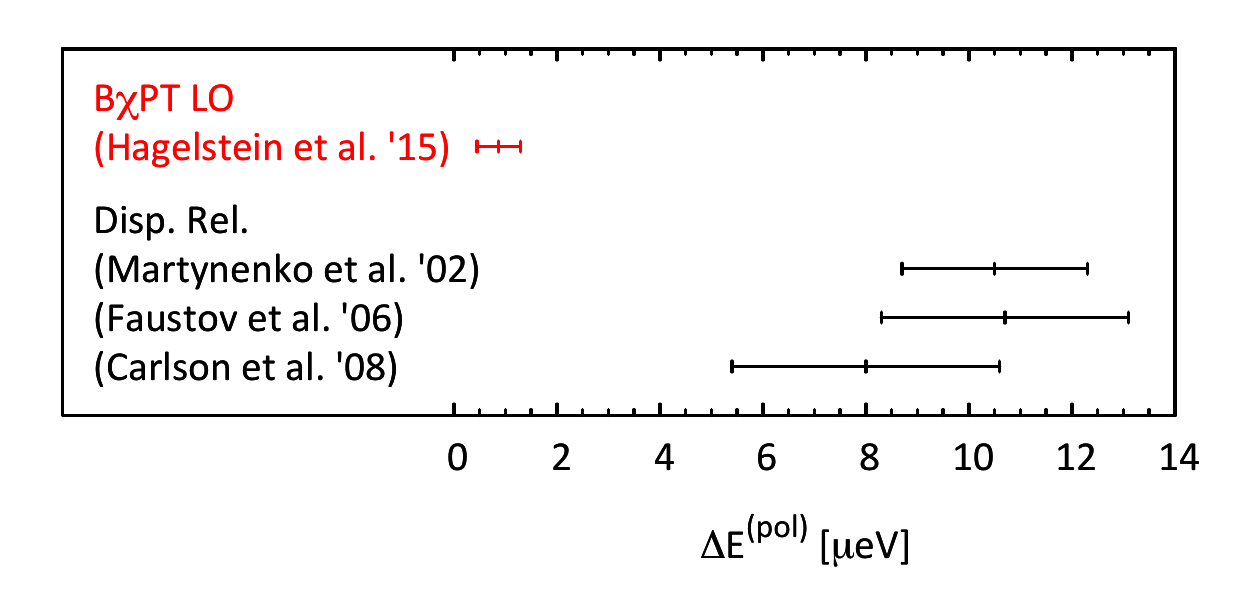}
\caption{Comparison of predictions for the $2S$ HFS in $\mu$H. The dispersive calculations are from Refs.~\cite{Carlson:2011af,Carlson:2008ke,Faustov:2006ve,Faustov:2001pn,Martynenko:2004bt}.}
\label{Comparison}
\end{figure}

Combining the individual pion contributions, Eqs.~\eref{PE} and \eref{PNLoop}, we arrive at:
\beq
E_{2S\,\mathrm{HFS}}^{(\pi^0 \,\&\,\pi N\,\mathrm{loops} )}=0.87\pm 0.42\, \upmu \mathrm{eV}.
\eeq
%\beq
%E_\mathrm{HFS}^{\pi^0 \,\&\,\pi N }=1.29\pm 0.26\, \upmu \mathrm{eV}.
%\eeq
Figure \ref{Comparison} shows a comparison between our BChPT prediction for the polarizability contribution to the $2S$ HFS in $\mu$H and other predictions based on dispersive approaches \cite{Carlson:2011af,Carlson:2008ke,Faustov:2006ve,Faustov:2001pn,Martynenko:2004bt}. Apparently, the two methods do not give consistent results; our prediction is considerably smaller.

%At first glance, the two methods do not give consistent results; our prediction is considerably smaller. Unfortunately, the interpretation is complicated by the fact that the usual definition of the polarizability contribution, used in dispersive calculations, contains also a term involving the elastic Pauli form factor:
%\beq
%E^{\mathrm{pol}(F_2)}_{2S \,\mathrm{HFS}}=E_F(2S) \frac{\al m}{2\pi(1+\varkappa) M} \int_0^\infty \frac{\dd Q^2}{Q^2} \beta_1(\tau_\mu) F_2^2(Q^2),\eqlab{F2}
%\eeq
%where $\tau_\mu=Q^2/4m^2$ and
%\beq
%\beta_1(\tau_\mu)=-3\tau_\mu+2\tau_\mu^2+2(2-\tau_\mu)\sqrt{\tau_\mu(\tau_\mu+1)}.
%\eeq
%For a proper comparison, one would need to know the $F_2$ contribution as estimated by the particular dispersive calculation. In the region of $Q^2\in [0.0452,20]\, \mathrm{GeV}^2$ Carlson et al.\ find $E^{\mathrm{pol}(F_2)}_{2S \,\mathrm{HFS}}\approx 7.23\pm0.23 \, \upmu\mathrm{eV}$ \cite{Carlson:2008ke} [cf.\ Table IV therein], what dominates their result for the polarizability contribution, cf.\ Fig.~\ref{Comparison}. However, evaluating the full integral \eref{F2} with the form factor parametrization of Bradford et al.~\cite{Bradford:2006yz} yields: $E^{\mathrm{pol}(F_2)}_{2S \,\mathrm{HFS}}=24.56 \, \upmu\mathrm{eV}$. 

However, one should mention that the LO BChPT prediction for the Lamb shift \cite{Alarcon:2013cba} is in good agreement with dispersive calculations, e.g., \cite{Carlson:2011zd,Birse:2012eb}. Despite the fact that the Lamb shift calculations involve a so-called ``subtraction term'', corresponding to the subtraction in the $T_1$ dispersion relation, which can only be modeled within the dispersive approaches.

%\section{Outlook and Conclusions}

For future work, one has to investigate if there are higher-order corrections contributing non-negligibly to the BChPT prediction, e.g., through the $\Delta$-excitation.


\begin{thebibliography}{99}


%\bibitem{Mohr:2012tt} 
 % P.~J.~Mohr, B.~N.~Taylor and D.~B.~Newell, {\it CODATA recommended values of the fundamental physical constants: 2010},  Rev.\ Mod.\ Phys.\  {\bf 84} (2012) 1527 [atom-ph/1203.5425].
  %%CITATION = ARXIV:1203.5425;%%
  
%  \bibitem{Bernauer:2010wm} 
 % J.~C.~Bernauer, et al.\ [A1 Collaboration], {\it High-precision determination of the electric and magnetic form factors of the proton}, Phys.\ Rev.\ Lett.\  {\bf 105} (2010) 242001 [nucl-ex/1007.5076]; 
  %%CITATION = ARXIV:1007.5076;%%
  %\cite{Bernauer:2013tpr}
%\bibitem{Bernauer:2013tpr} 
  %J.~C.~Bernauer {\it et al.}  [A1 Collaboration],
  %``The electric and magnetic form factors of the proton,''
% Phys.\ Rev.\ C {\bf 90} (2014) 015206 [nucl-ex/1307.6227].
 % arXiv:1307.6227 [nucl-ex].
  %%CITATION = ARXIV:1307.6227;%%
  
  %\cite{Zhan:2011ji}
%\bibitem{Zhan:2011ji} 
%  X.~Zhan, K.~Allada, D.~S.~Armstrong, J.~Arrington, W.~Bertozzi, W.~Boeglin, J.-P.~Chen, K.~Chirapatpimol, et al., {\it High precision measurement of the proton elastic form factor ratio $\mu_pG_E/G_M$ at low $Q^2$},
%  Phys.\ Lett.\ B {\bf 705} (2011) 59 [nucl-ex/1102.0318].
%  [arXiv:1102.0318 [nucl-ex/1102.0318]].
  %%CITATION = ARXIV:1102.0318;%%
  %87 citations counted in INSPIRE as of 04 Feb 2015
  
   %\citep{Antognini:1900ns}
\bibitem{Antognini:1900ns} 
  A.~Antognini, F.~Nez, K.~Schuhmann, F.~D.~Amaro, F.~Biraben, J.~M.~R.~Cardoso, D.~S.~Covita, A.~Dax, et al., {\it Proton structure from the measurement of $2S-2P$ transition frequencies of muonic hydrogen},
  Science {\bf 339} (2013) 417.
  %%CITATION = SCIEA,339,417;%%
  
   %\cite{Antognini:2013rsa}
\bibitem{Antognini:2013rsa} 
  A.~Antognini, F.~Kottmann, F.~Biraben, P.~Indelicato, F.~Nez and R.~Pohl, {\it Theory of the 2S-2P Lamb shift and 2S hyperfine splitting in muonic hydrogen}, Annals Phys.\  {\bf 331} (2013) 127 [atom-ph/1208.2637].
  %%CITATION = APNYA,331,127;%%
  
  
\bibitem{Hagelstein:2015yma}
F.~Hagelstein and V.~Pascalutsa, {\it Breakdown of the expansion of finite-size corrections to the hydrogen Lamb shift in moments of charge distribution}, Phys.\ Rev.\ A {\bf 91} (2015) 040502 [hep-ph/1502.03721].


\bibitem{Carlson:2011af}
C.~E.~Carlson, V.~Nazaryan and K.~Griffioen, {\it Proton structure corrections to hyperfine splitting in muonic hydrogen},
Phys.\ Rev.\ A {\bf 83} (2011) 042509 [atom-ph/1101.3239].

\bibitem{Alarcon:2013cba}
J.~M.~Alarcón, V.~Lensky and V.~Pascalutsa, {\it Chiral perturbation theory of muonic hydrogen Lamb shift: polarizability contribution}, Eur.\ Phys.\ J.\ C {\bf 74} (2014) 2852 [hep-ph/1312.1219]. 

\bibitem{Carlson:2008ke}
C.~E.~Carlson, V.~Nazaryan and K.~Griffioen, {\it Proton structure corrections to electronic and muonic
                        hydrogen hyperfine splitting},
Phys.\ Rev.\ A {\bf 78} (2008) 022517 [atom-ph/0805.2603].

\bibitem{Faustov:2006ve}
R.~N.~Faustov, I.~V.~Gorbacheva and A.~P.~Martynenko, {\it Proton polarizability effect in the hyperfine splitting of the hydrogen atom}, Proc.\ SPIE Int.\ Soc.\ Opt.\ Eng.\ {\bf 6165} (2006) 0M [hep-ph/0610332].

\bibitem{Faustov:2001pn}
R.~N.~Faustov, E.~V.~Cherednikova and A.~P.~Martynenko, {\it Proton polarizability contribution to the hyperfine splitting in muonic hydrogen}, Nucl.\ Phys.\ A {\bf 703} (2002) 365-377 [hep-ph/0108044].

\bibitem{Martynenko:2004bt}
A.~P.~Martynenko, {\it 2S hyperfine splitting of muonic hydrogen}, Phys.\ Rev.\ A {\bf 71} (2005) 022506 [hep-ph/0409107].

\bibitem{Zhou:2015bea}
H.-Q.~Zhou, H.-R.~Pang, {\it One-pion-exchange effect in the energy spectrum of muonic hydrogen}, Phys.\ Rev.\ A {\bf 92} (2015) 032512.


\bibitem{Bradford:2006yz}
R.~Bradford, A.~Bodek, H.~S.~Budd and J.~Arrington, {\it A new parameterization of the nucleon elastic form factors}, Nucl.\ Phys.\ B (Proc.\ Suppl.) {\bf 159} (2006) 127-132 [hep-ex/0602017].


\bibitem{Carlson:2011zd}
C.~E.~Carlson, M.~Vanderhaeghen, {\it Higher order proton structure corrections to the Lamb shift in muonic hydrogen}, Phys.\ Rev.\ A {\bf 84} (2011) 020102 [hep-ph/1101.5965].

\bibitem{Birse:2012eb}
M.~C.~Birse and J.~A.~McGovern, {\it Proton polarisability contribution to the Lamb shift in muonic hydrogen at fourth order in chiral perturbation theory}, Eur.\ Phys.\ J.\ A {\bf 48} (2012) 120 [hep-ph/1206.3030].

\end{thebibliography}
\end{document}